\documentclass[aps,prl,reprint,superscriptaddress]{revtex4-2}

\usepackage{bm}
\usepackage{amsmath}
\usepackage[dvipdfmx]{graphicx}
\usepackage{float}
\usepackage{color}

\usepackage {ulem}
\usepackage{cancel}

\graphicspath{{../Figure/}}

\begin{document}


\title{Korringa factor in Dirac-nodal-line systems}


\author{Takahiko Sekine}
\affiliation{Department of Applied Physics, University of Tokyo, Bunkyo-ku, Tokyo 113-8656, Japan}
\affiliation{Max-Planck-Institute for Solid State Research, Heisenbergstraße 1, Stuttgart 70569, Germany}
\author{Kazushi Kanoda}
\affiliation{Department of Applied Physics, University of Tokyo, Bunkyo-ku, Tokyo 113-8656, Japan}
\affiliation{Max-Planck-Institute for Solid State Research, Heisenbergstraße 1, Stuttgart 70569, Germany}
\affiliation{Physics Institute, University of Stuttgart, Pfaffenwaldring 57, 70550 Stuttgart, Germany}

\date{\today}

\begin{abstract}
The Korringa factor $K(\alpha)$ in the nuclear magnetic resonance studies is a widely used measure of electron-correlation effects in metals and is conventionally interpreted using the non-correlated value $K(\alpha)$ = 1. In Dirac-fermion systems, however, the non-correlated reference value depends on the electronic density of states (DOS) and differs from that of conventional metals. Here, we theoretically investigate the non-correlated Korringa factor $K^*(\alpha)$ in a Dirac nodal-line (DNL) system with finite nodal-line dispersion using a DOS-based approach. We show that $K^*(\alpha)$ exhibits an intrinsic temperature dependence, continuously evolving from the Dirac-fermion value of 1.71 at high temperatures to the metallic value of 1 at low temperatures. The crossover temperature is found to be proportional to the nodal-line dispersion width, identified as the characteristic energy scale governing the non-correlated Korringa factor in DNL systems. We further demonstrate that Landau quantization generates singular DOS structures that strongly enhance $K^*(\alpha)$, producing pronounced peaks even in the absence of electron correlations. In addition, $K^*(\alpha)$ exhibits two distinct types of magnetic oscillations arising from intra-level and inter-level contributions. The latter originates from the overlap of singular points belonging to positive and negative quantum-number Landau subbands at the Fermi level and represents a unique consequence of the finite energy dispersion of a DNL. These results provide a quantitative reference for interpreting NMR experiments in DNL materials and for identifying genuine correlation effects beyond DOS-induced contributions.
\end{abstract}


\maketitle

\section{I. Introduction}
Interactions among electrons in solids give rise to a variety of emergent phenomena, including charge ordering, magnetic ordering, metal–insulator transitions, and unconventional superconductivity \cite{Imada1998,
 Dagotto2005}. Even in paramagnetic metallic states, electron correlations can strongly modify low-energy spin excitations through the enhancement of spin fluctuations. Nuclear magnetic resonance (NMR) is one of the most powerful microscopic probes of such correlation effects because it directly accesses both static and dynamical spin susceptibilities by measuring the Knight shift, $K$, and the nuclear spin-lattice relaxation rate, $1/T_1$ \cite{Slichter1990}. In metallic systems, $K$ and $1/T_1$ are related through the Korringa relation \cite{Moriya1963, Korringa1950},
\begin{equation}
\frac{1}{T_1TK^2} = \frac{4\pi k_{\rm B}}{\hbar}\beta K(\alpha),
\end{equation}
where $T$ is temperature, $\hbar$ is the reduced Planck constant, $k_{\rm B}$ is the Boltzmann constant, and $\beta$ is the square of the ratio between the transverse and longitudinal components of the hyperfine-coupling tensor. The dimensionless quantity $K(\alpha)$, known as the Korringa factor, provides a measure of electron-correlation effects. In a non-correlated electron gas, $K(\alpha) = 1$. Antiferromagnetic spin correlations generally enhance $1/T_1T$, leading to $K(\alpha) > 1$, whereas ferromagnetic correlations result in $K(\alpha) < 1$. Consequently, the Korringa factor has been widely used as a quantitative indicator of spin correlations in metallic systems.

However, the derivation of this Korringa relation assumes that the electronic density of states (DOS) varies only weakly within an energy window of kBT order around the chemical potential. This assumption breaks down in gapless Dirac and Weyl systems, where the DOS exhibits a strong energy dependence near the band-touching point \cite{Dora2009, Hirata2017, Hirata2021}. As a result, the non-correlated value of the Korringa factor is no longer universal.
A representative example is a two-dimensional (2D) massless Dirac fermion (DF) system. When the chemical potential is located at the Dirac point, the non-correlated Korringa factor takes the value $K(\alpha) = \pi(\ln 2)^2/12 \sim 1.71$ instead of $K(\alpha)=1$ \cite{Hirata2017}. Therefore, $K(\alpha)>1.71 (<1.71)$ indicates antiferromagnetic (ferromagnetic) spin correlations. This concept has been successfully applied to NMR studies of the organic DF system $\alpha$-(BEDT-TTF)$_2$I$_3$, where weak antiferromagnetic correlations were identified at room temperature and strong interaction effects emerged at low temperatures \cite{Hirata2017, Hirata2021, Ohki2020}.
These studies demonstrated that a reliable determination of the non-correlated reference value $K^*(\alpha)$ is indispensable for extracting genuine correlation effects from NMR measurements in DF materials.

Dirac-nodal-line (DNL) semimetals provide a unique platform in which metallic and DF characteristics coexist. In a DNL system, Dirac points are continuously connected in momentum space to form a line node. In many realistic DNL materials, the nodal line possesses a finite energy dispersion. When the chemical potential lies within this dispersion, small electron and hole pockets emerge around the Fermi energy $E_{\rm F}$, giving rise to a finite DOS at $E_{\rm F}$. Consequently, the low-energy electronic properties are metallic, whereas DF characteristics dominate at higher energies outside the nodal-line dispersion range. Such a coexistence naturally raises the question of how the non-correlated Korringa factor $K^*(\alpha)$ evolves between the metallic limit $K^*(\alpha) = 1$ and the DF limit $K^*(\alpha) =1.71$. Unlike conventional metals or ideal 2D DF systems, the non-correlated Korringa factor in a DNL system is therefore expected to exhibit an intrinsic temperature dependence.

A number of DNL materials have been reported, including ZrSiX (X = S, Te) \cite{Topp2016,Tian2021,Fu2019}, ZrX$_2$ (X = Se, Te) \cite{Kar2020, Tian2020}, AlB$_2$ \cite{Takane2018}, CaAgX (X = P, As) \cite{Hirose2020, Yamakage2016, Takane2018a}, [Pd(dddt)$_2$] \cite{Kato2017, Kato2017a}, and [M(dmdt)$_2$] (M = Ni, Pt) \cite{Zhou2019, Kobayashi2021, Kawamura2020}. In the molecular conductors [Pd(dddt)$_2$] and [M(dmdt)$_2$], the electronic states near the Fermi level are predominantly derived from DNL bands without contributions from trivial metallic bands.
Thus, in order to investigate the electron correlation effect in DNL through NMR, determining the reference value of $K^*(\alpha)$ in DNL materials is of practical importance.

Another important aspect is the effect of magnetic fields. NMR measurements are necessarily performed under finite magnetic fields, and DF exhibit unconventional Landau quantization.
Especially in the DNL system, the quantized Landau level can be dispersive along the DNL and form subbands. The resulting Landau subbands generate characteristic singular structures in the DOS, which may substantially modify the Korringa factor.
Therefore, understanding the magnetic-field dependence of $K^*(\alpha)$ is essential for interpreting NMR experiments on DNL materials.

In the present study, we calculate the non-correlated Korringa factor $K^*(\alpha)$ for a DNL system using the DOS approximation.
We first investigate the zero-field limit and demonstrate a temperature-driven crossover between the DF and metallic regimes. We then examine the influence of Landau quantization under magnetic fields and clarify how singular DOS structures modify $K^*(\alpha)$.
The results provide a quantitative reference for future NMR studies of DNL materials and establish a basis for identifying genuine correlation effects in nodal-line systems.

\section{II. Method}
\subsection{A. DOS approximation}
\begin{figure}
\centering
\includegraphics[width=.8\columnwidth]{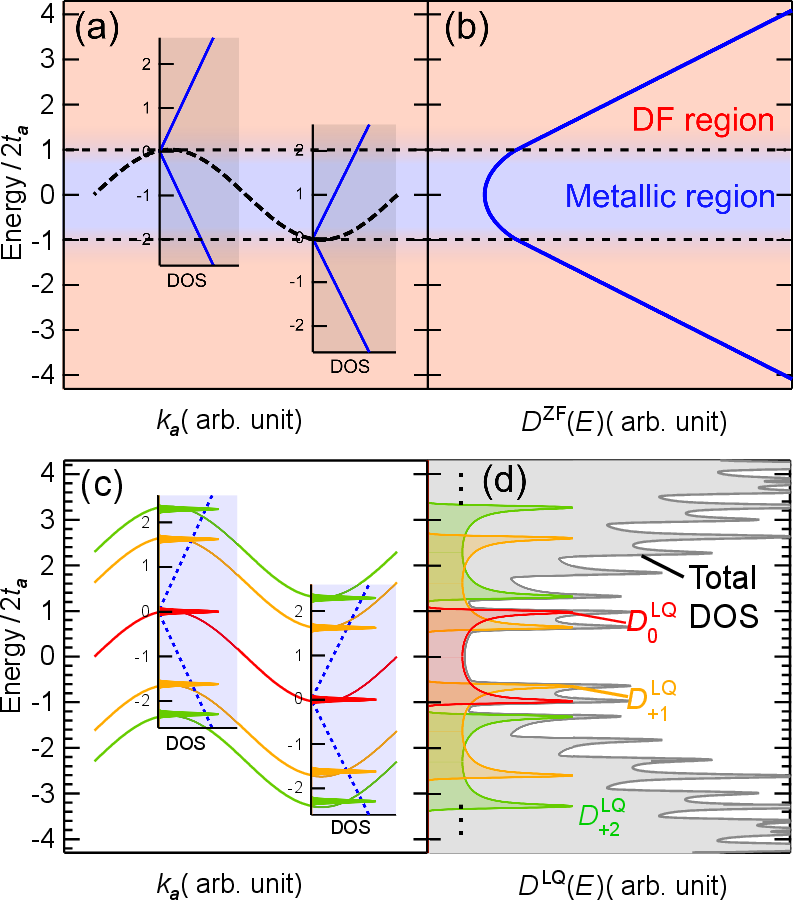}
\caption{\label{Fig1}(Color online) Dispersion and density of states of DNL with and without Landau quantization used in the simulations. (a) $k_a$ dispersion of the 2D DF DOS and (b) the entire DOS in the DNLs under zero field. (c) $k_a$ dispersions of the Landau levels in the 2D plane and (d) the DOS of the DNL under a finite field. (d) the DOS of each Landau level distinguished by different colors and the total DOS shown by a gray region.}
\end{figure}

In this study, we evaluate the non-correlated Korringa factor $K^*(\alpha)$ within a DOS approximation, neglecting electron-correlation effects throughout. The Knight shift K is proportional to the uniform static spin susceptibility and is expressed as
\begin{equation}
K=\frac{\mu_{\rm B}A_\parallel}{2}\int dED(E)\left( -\frac{\partial f}{\partial E} \right),
\end{equation}
where $D(E)$ is the electronic DOS, $\mu_{\rm B}$ is the Bohr magneton, and $f(E)$ is the Fermi–Dirac distribution function. The chemical potential is fixed at the charge-neutral point throughout this work. $1/T_1T$ is related to the imaginary part of the transverse dynamical spin susceptibility as \cite{Moriya1963,Narath1968}
\begin{equation}
\frac{1}{T_1T}=\frac{2\gamma_{\rm n}^2 k_{\rm B}A_\perp^2}{\gamma_{\rm e}^2}
\sum_{\bm{q}}
\frac{{\rm Im}\chi_\perp(\bm{q},\omega_{\rm N})}{\omega_{\rm N}}
\end{equation}
where $\gamma_{\rm e}$ and $\gamma_{\rm n}$ are the electron and nuclear gyromagnetic ratios, respectively, and $\omega_{\rm N}$ is the NMR frequency. For a non-interacting electron system, the low-energy limit of Eq. (3) can be expressed in terms of the DOS as \cite{Moriya1963}
\begin{eqnarray}
\frac{1}{T_1T} &=& \frac{\pi}{T\hbar} (\gamma_{\rm e} \gamma_{\rm n} \hbar^2)^2A_\perp^2\notag \\
 &&\times\iint dEdE' D_\uparrow(E)D_\downarrow(E')f(E) \left\{1-f(E')\right\} \delta (E-E')\notag \\
&\simeq& \frac{\pi k_{\rm B}}{\hbar} (\gamma_{\rm e} \gamma_{\rm n} \hbar^2)^2A_\perp^2 \int dE\left\{D(E)\right\}^2 \left(-\frac{\partial f}{\partial E}\right)
\end{eqnarray}
where $D\uparrow(E)$ and $D\downarrow(E)$ denote the spin-resolved DOS. For simplicity, Zeeman splitting is neglected in the present calculations. Substituting Eqs. (2) and (4) into Eq. (1), we obtain the DOS-based expression of the Korringa factor,
\begin{equation}
K^*(\alpha) = \frac{\int dE \left\{D(E)\right\}^2 \left(-\frac{\partial f}{\partial E}\right)}{\left\{\int dE D(E) \left(-\frac{\partial f}{\partial E}\right)\right\}^2}.
\end{equation}
Thus, within the present framework, $K^*(\alpha)$ is determined solely by the DOS profile around the chemical potential. To evaluate Eq. (5), we construct model DOS functions based on simplified electronic dispersions of DNL systems.

\subsection{B. Zero-field DNL model}
We first consider a zero-field (ZF) model without Landau quantization. The energy dispersion of a DNL system is described by
\begin{equation}
E^{\rm ZF}_{\bm{k}} = \hbar v_{\rm F} \sqrt{k^2_x + k^2_y} + 2t_a \cos ak_z.
\end{equation}
The Fermi velocity $v_{\rm F}$ is assumed to be independent of wavenumbers for the non-tilting Dirac cones in this equation. $2t_a$ is half of the energy dispersion width of the DNL, and a is the lattice constant. $k_x$, $k_y$ and $k_z$ are the wavenumbers orthogonal to each other.
Equation (6) expresses that the Dirac points of the 2D Dirac cones on the $k_x$--$k_y$ planes (blue regions in Fig. 1(a)) are connected in lines along the $k_z$ axis (red line in Fig. 1(a)), representing an {\it open} DNL in the first Brillouin zone. The corresponding DOS, $\sum_{\bm{k}}\delta(E-E_{\bm{k}})$, is given by 
\begin{equation}
D^{\rm ZF}( E) = \dfrac{|E|}{2a\pi (\hbar v_{\rm F})^2}\qquad \qquad \left( |E| \geq 2t_a \right) \label{DOS_ZF1} \\
\end{equation}
and
\begin{align}
D^{\rm ZF}(E) &= \dfrac{|E|\left(\pi - 2 \cos ^{-1} [E/2t_a]\right)+4t_a \sqrt{1-(E/2t_a)^2}}{2a\pi^2 \left( \hbar v_{\rm F}\right)^2}\notag \\
&\hspace{4.6cm}\left( |E| < 2t_a\right). \label{DOS_ZF2}
\end{align}
As also seen in Fig. 1(b), the DOS exhibits two characteristic regimes. For $|E| > 2t_a$, the DOS increases linearly with energy, reflecting the 2D DF character (red region). In contrast, for $|E| < 2t_a$, the finite energy dispersion of the nodal line produces a nonzero DOS at the Fermi level, corresponding to small electron and hole pockets (blue region).
Consequently, the model naturally incorporates both metallic and DF characteristics.

\subsection{C. Landau-quantized DNL model}
To investigate magnetic-field effects, we next introduce Landau quantization of the Dirac cones. The resulting Landau-quantized subbands due to the DNL dispersion are labelled by an integer $N$ and described by
\begin{equation}
E^{\rm LQ}_N (k_z, H) = sgn(N) v_{\rm F} \sqrt{2e\hbar \mu_0 H|N|} + 2t_a \cos ak_z.
\end{equation}
The DOS of the $N$-th subband is
\begin{equation}
D^{\rm LQ}_N (E,H) = \frac{e\mu_0H}{2\pi \hbar} \sum_{k_z} \delta \left(E-E^{\rm LQ}_N (k_z, H) \right)
\end{equation}
and the total DOS is obtained by summing over all Landau indices $N$ (Fig. 1(d)). Each subband possesses two singular points (SPs) at the edges of the dispersion, where the DOS diverges.
To avoid divergences in the numerical integration of Eq. (5), a finite linewidth is introduced by convoluting the energy levels with a Gaussian broadening function.
Strictly speaking, a Landau-quantized system does not satisfy the assumptions underlying the original Korringa relation because the DOS varies strongly within a narrow energy range.
Nevertheless, Eq. (5) remains a useful quantity for characterizing how singular DOS structures affect the relationship between $K$ and $1/T_1T$. In the following, we therefore employ Eq. (5) as a DOS-sensitive measure of the generalized Korringa factor under magnetic fields.

\begin{figure}
\centering
\includegraphics[width = .8\columnwidth]{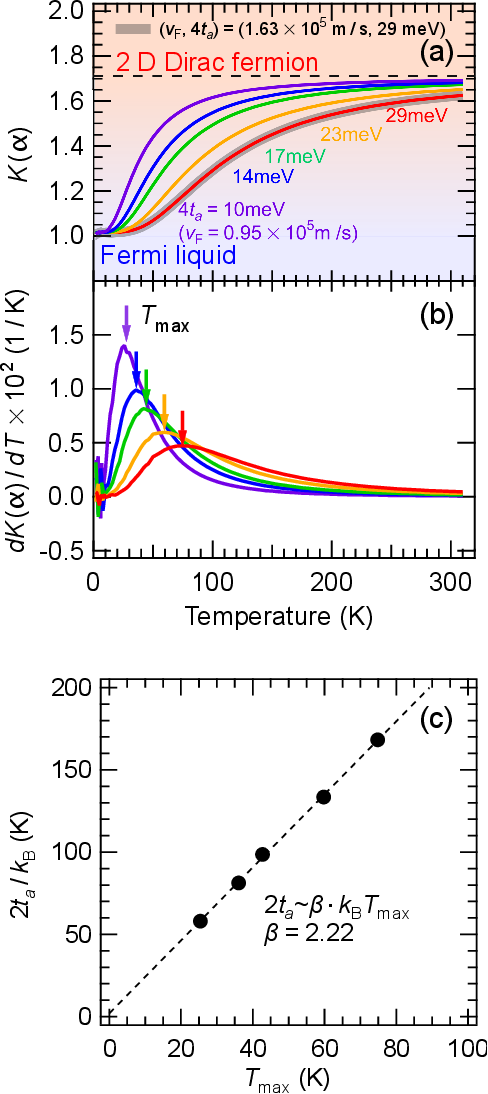}
\caption{\label{Fig2}(Color online) (a) $K^*(\alpha)$ calculated using the ZF model with $(v_{\rm F}/[10^5{\rm m/s}], 4t_a/[{\rm meV}]) = (1.63, 29)$ (bold gray line), $(0.95, 29)$(red), $(0.95, 23)$(orange), $(0.95, 17)$(green), $(0.95, 14)$(blue) and $(0.95, 10)$(purple). (b) Temperature derivative of $K^*(\alpha)$, $dK^*(\alpha)/dT$. The temperatures of the peaks, Tmax, are indicated by arrows. (c) $T_{\rm max}$ plotted against $2t_a$. The dotted line represents a fit to the linear function $2t_a=\beta k_{\rm B} T_{\rm max}$, where the temperature factor $\beta=2.22$.}
\end{figure}

\subsection{D. Model parameters}
The parameters $v_{\rm F}$ and $4t_a$ are varied independently to clarify their respective roles in determining $K^*(\alpha)$. As realistic reference values, we adopt parameter sets obtained by previous band-structure calculations of the molecular DNL conductor [Ni(dmdt)$_2$] \cite{Kobayashi2021, Kawamura2020, Kawamura2021}.
First-principles calculations yield $(v_{\rm F}, 4t_a) = (1.63\times10^5~{\rm m/s}, 29~{\rm meV})$.
{\color{black}In addition to these values, we also employed reduced $(v_{\rm F}, 4t_a)$ with considering bandwidth suppression by electron correlation effects.
Especially, $(v_{\rm F}, 4t_a) = (0.95\times10^5~{\rm m/s}, 17~{\rm meV)}$ are used as representative parameters for the Landau-quantized (LQ) model. They are approximately 60\% of the first-principles-calculation values, which is a reasonable a reduction in a real material.
The LQ model with those parameters successfully show specific magnetic-field effects in $K^*(\alpha)$ within the measurable field range for practical NMR experiments as discussed below, which do not rely on the specific properties of [Ni(dmdt)$_2$] but rather reflect generic consequences of a DNL with finite energy dispersion.}

\begin{figure*}
\centering
\includegraphics[width = 1.8\columnwidth]{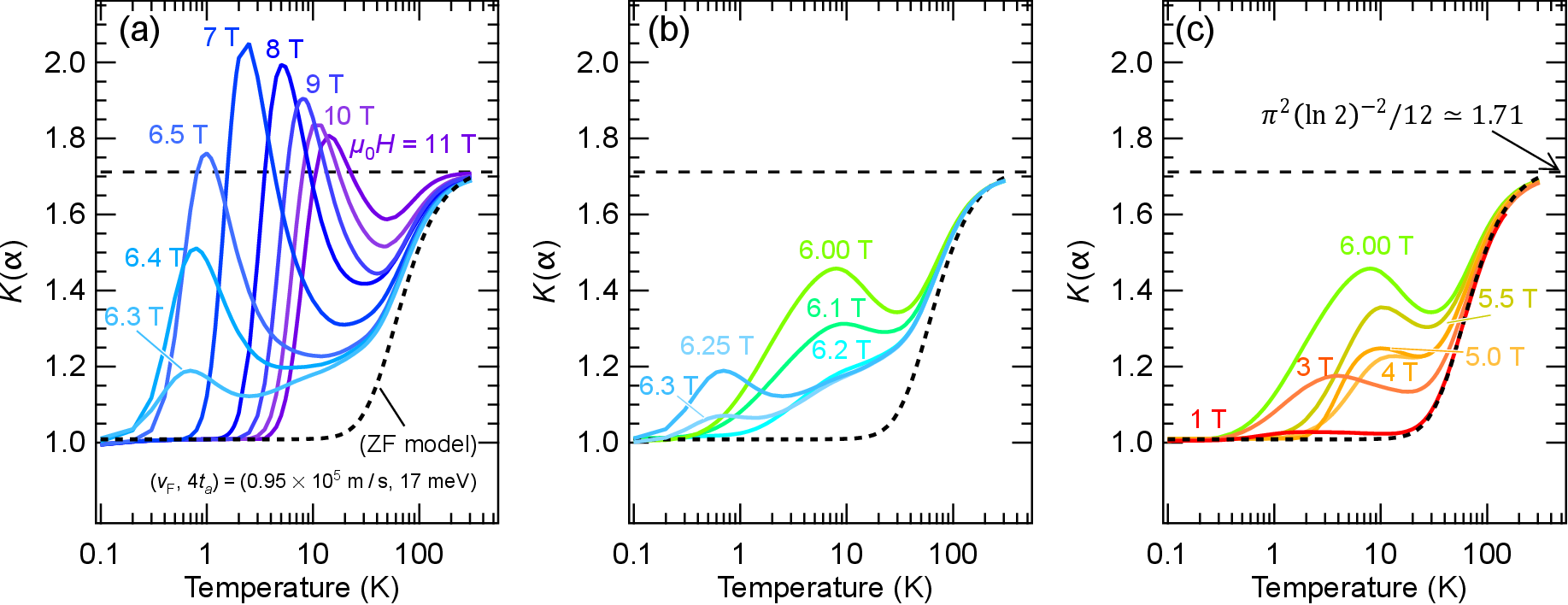}
\caption{\label{Fig3}(Color online) $K^*(\alpha)$ calculated using the LQ model under $\mu_0 H= 11$--6.3~T (a), 6.3--6.00~T (b), and 6.00--1.0~T (c) with $(v_{\rm F}/[10^5{\rm m/s}], 4t_a/[{\rm meV]}) = (0.95, 17)$. The ZF-model $K^*(\alpha)$ with the same $(v_{\rm F}, 4t_a)$ is represented by the black dotted curve (as shown in green FIG. 2(a)).}
\end{figure*}
\begin{figure*}
\centering
\includegraphics[width =1.8\columnwidth]{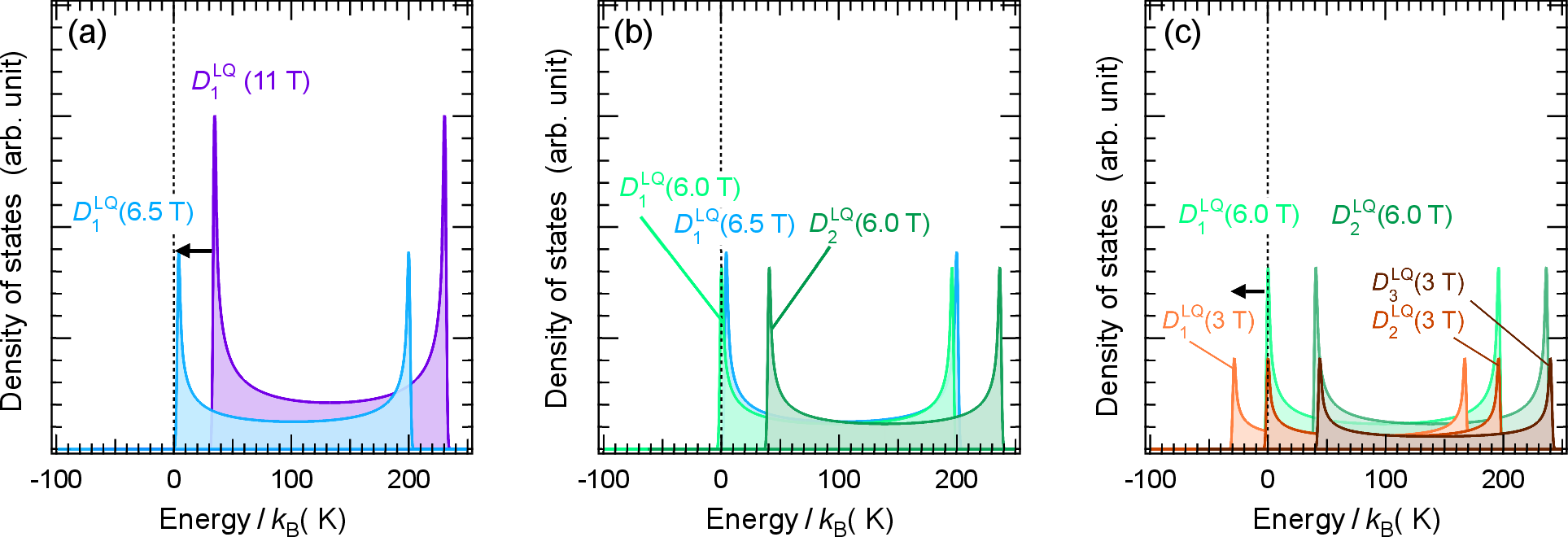}
\caption{\label{DOSsimulations}(Color online) Density of states of the Landau-quantized subbands in the DNL, $D_N^{\rm LQ}$, with the same parameters $v_{\rm F}$ and $4t_a$ as in Fig. 3. $D_1^{\rm LQ}$ under $\mu_0 H= 11$~T ((a)purple), $D_1^{\rm LQ}$ under 6.5~T ((a, b)light blue), $D_1^{\rm LQ}$ and $D_2^{\rm LQ}$ under 6.0~T ((b, c)light green and dark green, respectively), $D_1^{\rm LQ}$, $D_2^{\rm LQ}$ and $D_3^{\rm LQ}$ under 3.0~T ((c)orange, dark orange and brown, respectively). Here only the $N > 0$ terms of the $D_N^{\rm LQ}$ are shown; however, the $N \leq 0$ terms were also considered in the calculation of $K^*(\alpha)$ using Eq. 5.}
\end{figure*}

\section{III. Results and discussions}
\subsection{A. Temperature-dependent non-correlated Korringa factor}
The most fundamental consequence of a finite nodal-line dispersion is the emergence of a temperature-dependent non-correlated Korringa factor. Figure 2(a) shows $K^*(\alpha)$ calculated using the ZF model. For both parameter sets, $K^*(\alpha)$ approaches the DF value of 1.71 at high temperatures and continuously decreases toward the metallic value of 1 upon cooling.
This behavior reflects a crossover from a DF regime to a metallic regime in the DNL system. The origin of this crossover can be understood from the DOS profile shown in Fig. 1(b).
At high temperatures, the thermal window extends beyond the nodal-line dispersion width, and both K and $1/T_1T$ are predominantly influenced by the linear-in-energy DOS characteristic of a 2D DF.
 Consequently, $K^*(\alpha)$ approaches the DF limit of 1.71. Upon cooling, only the low-energy region around the Fermi level contributes to the NMR response. Because the finite nodal-line dispersion generates electron and hole pockets with a nonzero DOS at EF, the system gradually acquires metallic character and $K^*(\alpha)$ approaches unity.

A notable feature is that this crossover occurs even though the chemical potential remains fixed at the charge-neutral point. This behavior differs fundamentally from that of a conventional 2D DF system, where a crossover to metallic behavior requires a finite shift of the chemical potential away from the Dirac point. In the DNL system, the finite dispersion of the nodal line itself provides the energy scale responsible for the crossover. The crossover temperature strongly depends on the nodal-line dispersion width $4t_a$. As shown in Fig. 2(a), larger values of $4t_a$ shift the crossover toward higher temperatures. To quantify this behavior, we define the crossover temperature Tmax as the peak position of $dK^*(\alpha)/dT$ (Fig. 2(b)). Figure 2(c) demonstrates that Tmax is proportional to $4t_a$. This result indicates that the nodal-line dispersion width constitutes the sole characteristic energy scale governing the crossover. Remarkably, the Fermi velocity has no influence on the temperature profile of $K^*(\alpha)$ as evidenced by the exact correspondence of $K^*(\alpha)$ for $v_{\rm F} = 1.63\times10^5~{\rm m/s}$ (bold gray line) and $0.95\times10^5~{\rm m/s}$ (red line) with the fixed $4t_a$ value of 29~meV.

These results establish the first central conclusion of this work: Unlike ordinary metals or ideal DF systems, a DNL possesses an intrinsically temperature-dependent non-correlated Korringa factor that continuously connects the metallic limit $K^*(\alpha)=1$ and the DF limit $K^*(\alpha)=1.71$.

\begin{figure}
\centering
\includegraphics[width =0.75\columnwidth]{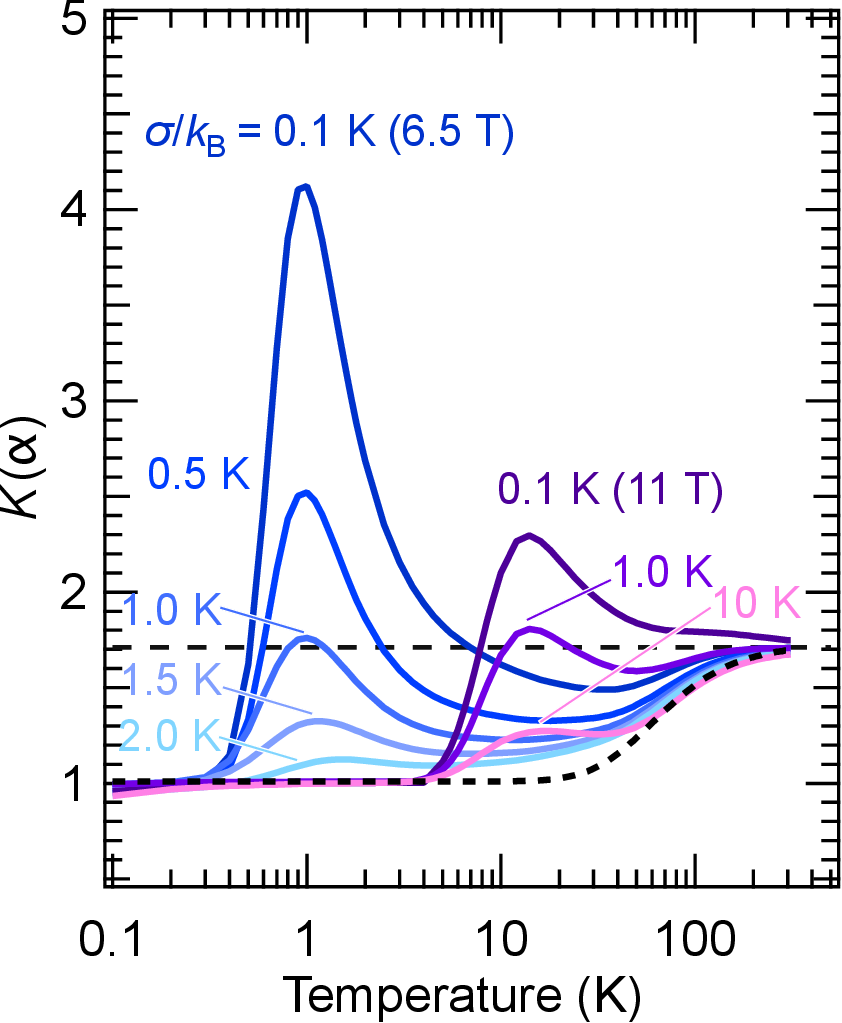}
\caption{\label{fig5}(Color online) LQ-model $K^*(\alpha)$ with various linewidths $\sigma$ under $\mu_0 H= 11$~T and 6.5~T.
$\sigma/k_{\rm B}$ is varied from 2.0~K (light blue) to 0.1~K (dark blue) under 6.5~T and from 10~K (pink) to 0.1~K (dark purple) under 11~T.}
\end{figure}

\subsection{B. Landau-quantization-induced enhancement of $K^*(\alpha)$}
We next investigate the effect of Landau quantization on the non-correlated Korringa factor. Figure 3 shows $K^*(\alpha)$ calculated using the LQ model.
In contrast to the smooth crossover observed in the zero-field model, $K^*(\alpha)$ exhibits pronounced peaks under magnetic fields. The peak temperature and amplitude depend strongly on the magnetic field and the Landau-level linewidth.
The origin of these peaks is the singular DOS generated at the edges of the Landau-quantized DNL subbands. As shown in Fig. 4, each subband possesses two singular points where the DOS diverges.
From Eq. (9), this divergence is similar to that of DOS at edges of a 1D metallic band, whose energy dependence can be approximated by $1/\sqrt{E-E_N\pm2t_a}$ near the SPs. Thus, K, the integration of the DOS itself (Eq. (2)), converges to a finite value even with zero linewidth of each Landau level, but $1/T_1T$, the integration of the square of the DOS (Eq. (4)), becomes divergently enhanced. Eventually, these singularities enhance $1/T_1TK^2$ more efficiently than $K$, resulting in a pronounced increase in $K^*(\alpha)$.

Under a magnetic field of 11~T, $K^*(\alpha)$ exhibits a large peak around 15 K (Fig. 3(a)). This peak originates from the SPs of the $N = \pm1$ Landau subband (Fig. 4(a)).
As the magnetic field decreases, the SPs move closer to EF because the Landau-level energy scales as $\sqrt{H}$.  Consequently, the peak shifts toward lower temperatures.
The peak amplitude initially increases because lower temperatures provide a sharper thermal weighting through the derivative of the Fermi--Dirac distribution function.
However, at still lower fields the peak is suppressed owing to the reduced Landau-level degeneracy and the finite linewidth broadening.

Under lower magnetic fields, multiple Landau-quantized subbands contribute simultaneously to $K^*(\alpha)$.
As shown in Fig. 3(b), the prominent low-temperature peak associated with the SPs of the $N = \pm1$ subbands is gradually suppressed as its characteristic energy approaches the Fermi level and becomes comparable to the Landau-level linewidth.
At the same time, an additional peak emerges around 10~K.
This second peak originates from the SPs of the $N = \pm2$ subbands, which enter the experimentally relevant temperature range below approximately 6~T, as illustrated in Fig. 4(b).
Upon further decreasing the magnetic field (Fig. 3(c)), higher-index subbands such as $N = \pm3$ also approach the $E_{\rm F}$ and contribute to $K^*(\alpha)$.
Consequently, the Korringa factor is no longer dominated by a single SP but rather by a collection of weaker singularities originating from multiple Landau levels. Because the degeneracy of each subband decreases with decreasing magnetic field, the contribution from an individual SP becomes smaller.
The resulting $K^*(\alpha)$ therefore evolves from a sharp single peak at high fields into a broader and weaker structure at lower fields.

Figure 5 further demonstrates that the peak amplitude in $K^*(\alpha)$ is strongly affected by the linewidth broadening. Narrower linewidths preserve the singular DOS structure and substantially enhance $K^*(\alpha)$, whereas broader linewidths smear the singularities and suppress the peak. The suppression is particularly effective when the characteristic temperature associated with the SP approaches the linewidth scale.

It should be emphasized that the enhancement of $K^*(\alpha)$ discussed above does not originate from electron correlations.
All calculations are performed within the DOS approximation and therefore represent a non-correlated reference value. The peak arises solely from singular DOS structures generated by Landau quantization.
Consequently, an enhanced $K^*(\alpha)$ under magnetic fields should not automatically be interpreted as evidence of antiferromagnetic spin fluctuations.

\begin{figure*}
\centering
\includegraphics[width =1.8\columnwidth]{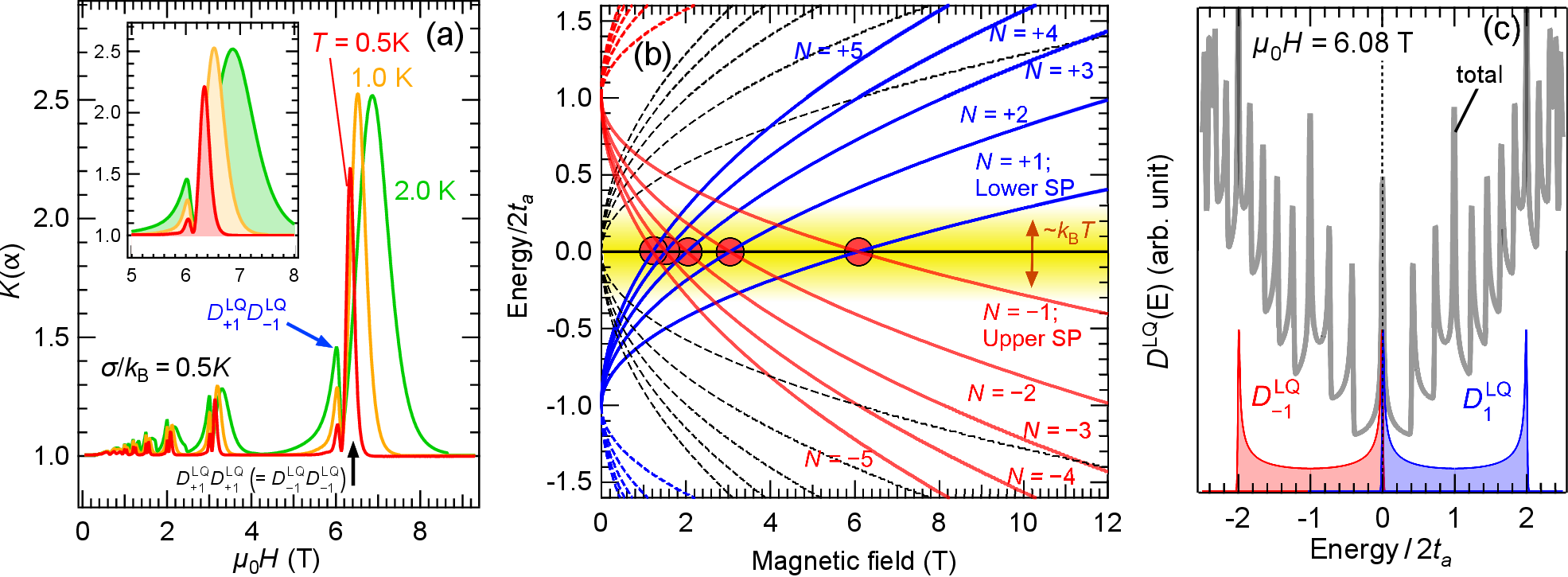}
\caption{\label{Fig6}(Color online) (a) Magnetic dependence of the LQ-model $K^*(\alpha)$ with $\sigma/k_{\rm B} = 0.5$~K at temperatures of 2.0~K (green), 1.0~K (yellow) and 0.5~K (red). The black arrow signifies the temperature-dependent peak provided by $D_{+1}^{\rm LQ}D_{+1}^{\rm LQ}$ and $D_{-1}^{\rm LQ}D_{-1}^{\rm LQ}$. The blue arrow signifies the temperature-independent peak provided by $D_{+1}^{\rm LQ}D_{-1}^{\rm LQ}$ and $D_{-1}^{\rm LQ}D_{+1}^{\rm LQ}$. The two peaks around 6~T are magnified in the inset. (b) Magnetic-field dependence of the Landau levels (black dotted), the upper SPs of $N < 0$ Landau subbands (red) and lower SPs of $N > 0$ Landau subbands (blue), respectively. The crossing points at $E_{\rm F}$ are marked by red circles. (c) Total DOS under $\mu_0 H= 6.08$~T (gray) where the upper SP of $N = -1$ and the lower SP of $N = +1$ coincide. The parts of $D_{+1}^{\rm LQ}$ and $D_{-1}^{\rm LQ}$ are highlighted by blue and red areas, respectively.}
\end{figure*}

Nevertheless, the singular DOS generated by Landau quantization may provide favorable conditions for correlation-driven instabilities. The quantized DNL subbands possess effectively 1D dispersions, and low-dimensional electronic systems are well known to exhibit enhanced tendencies toward spin-density-wave and superconducting instabilities \cite{Dressel2007, Barthel1993}.
Therefore, although the $K^*(\alpha)$ peak itself is non-correlated in origin, it suggests the possibility that interaction effects should become particularly important in real DNL materials under a certain magnetic field.

\subsection{C. Magnetic-field oscillations of $K^*(\alpha)$}
Finally, we examine the magnetic-field dependence of the non-correlated Korringa factor. Figure 6(a) shows $K^*(\alpha)$ calculated as a function of a magnetic field at several temperatures.
The oscillatory behavior exhibits two distinct components: a temperature-independent peak indicated by the blue arrow and a temperature-dependent peak indicated by the black arrow.
As discussed below, these two oscillations originate from different scattering processes and reflect a unique consequence of the Landau-quantized DNL electronic structure.

As the magnetic field varies, the SPs associated with the Landau-quantized DNL subbands move relative to $E_{\rm F}$ (Fig. 6(b)).  A particularly strong enhancement occurs when the lower SPs originating from the $N = +n$ and the upper SPs from the $N = -n$ subbands overlap at $E_{\rm F}$as highlighted by red circles in Fig. 6(b). This condition is generally expressed as $sgn(N) v_{\rm F} \sqrt{2e\hbar\mu_0H|N|}=2t_a$. For example, $\mu_0 H=6.08$~T satisfies this condition for $N = \pm1$ in Fig. 6(c), where the overlap of the upper SP of $N = -1$ and the lower SP of $N = +1$ makes the singularity in DOS more prominent at $E_{\rm F}$.
Because this overlap condition is determined solely by the electronic structure, the corresponding peak position is independent of temperature.
The overlap of the two singularities produces an exceptionally large contribution to $K^*(\alpha)$ through the inter-level terms $D_N\times D_{-N}$, which represent scattering processes between the conduction-band $(N > 0)$ and valence-band $(N < 0)$ Landau subbands.

The overlapping SPs separate and move away from $E_{\rm F}$ as the magnetic field increases further.
Consequently, the inter-level contribution $D_N\times D_{-N}$ is reduced, whereas the intra-level terms $D_N\times D_{N}$ become relatively more dominant.
This $D_N\times D_{N}$ terms can contribute to $K^*(\alpha)$ to some extent as far as the SPs are located within the thermal window of width $\sim k_{\rm B}T$ centered on $E_{\rm F}$(Fig. 6(b)).
The magnitude of such contributions should be enhanced by a magnetic field because Landau degeneracy itself is also enhanced (Eq. (9)).
In this regime, $K^*(\alpha)$ exhibits an additional peak, indicated by the black arrow in Fig. 6(a).
This peak occurs when the SPs of an individual Landau subband move out of the thermal window. Because the thermal window broadens with increasing temperature, the corresponding peak position shifts toward higher magnetic fields. The resulting oscillation is therefore temperature-dependent.

These two mechanisms occur repeatedly across successive Landau subbands and, collectively, give rise to the oscillatory field dependence of $K^*(\alpha)$.
In particular, the temperature-independent oscillation is a remarkable consequence of the finite energy dispersion of the DNL.
In a conventional 2D DF system, positive and negative Landau levels remain separated in energy, preventing their DOS singularities from overlapping at $E_{\rm F}$.
In contrast, the finite nodal-line dispersion broadens each Landau level into a dispersive subband with SPs at the edges of the DOS, allowing the SPs of the $N > 0$ and $N < 0$ branches to overlap.
The resulting inter-level contribution therefore produces an oscillation absolutely unique to dispersive DNL systems.

These results suggest that magnetic-field-dependent measurements of quantities governed primarily by the DOS, such as magnetic susceptibility or electrical resistivity, may exhibit signatures analogous to those shown in Fig. 6.
In particular, the coexistence of conventional intra-level oscillations and DNL-specific inter-level oscillations could serve as an experimental hallmark of dispersive DNLs.

\section{IV. Conclusion}
In this study, we investigated the non-correlated Korringa factor $K^*(\alpha)$ in a DNL system within a DOS approximation. Unlike conventional metals and ideal 2D DF systems, a DNL with finite nodal-line dispersion exhibits an intrinsically temperature-dependent Korringa factor.
$K^*(\alpha)$ continuously evolves from the DF value of 1.71 at high temperatures to the metallic value of 1 at low temperatures, reflecting a crossover between DF and metallic regimes.
We demonstrated that the characteristic crossover temperature is proportional to the nodal-line dispersion width $4t_a$, establishing the latter as the fundamental energy scale governing the non-correlated Korringa factor in DNL systems.

We further investigated the influence of Landau quantization on $K^*(\alpha)$.
The singular DOS structures at the edges of Landau-quantized DNL subbands strongly enhance $K^*(\alpha)$, producing pronounced peaks that originate entirely from non-correlated electronic states.
These results provide an important reference for distinguishing purely DOS-induced enhancements from genuine spin-correlation effects in NMR experiments.

Finally, we showed that $K^*(\alpha)$ exhibits two distinct types of magnetic-field oscillations.
One originates from intra-level contributions and exhibits a characteristic temperature dependence, whereas the other arises from inter-level contributions associated with the overlap of SPs belonging to positive and negative Landau subbands.
The latter oscillation is a unique consequence of the finite energy dispersion of a DNL and has no counterpart in conventional 2D DF systems.

The present results establish a quantitative framework for interpreting Korringa factors in DNL materials and provide a non-correlated reference for future NMR investigations of electron-correlation effects in nodal-line systems.

\section{Acknowledgement}
This work  was supported by Japan Society for the Promotion of Science (21K18144 and Core-to-Core Program (JPJSCCA20240001)) and by the Mitsubishi Foundation (Grant Number 202110014). K. K. acknowledges the support provided by the funding from the Alexander von Humboldt Foundation.
\bibliography{Korringa.bib}

@article{Korringa1950,
author = {Korringa, J.},
doi = {10.1016/0031-8914(50)90105-4},
issn = {00318914},
journal = {Physica},
month = {jul},
number = {7-8},
pages = {601--610},
url = {https://linkinghub.elsevier.com/retrieve/pii/0031891450901054},
volume = {16},
year = {1950}
}

@article{Topp2016,
author = {Topp, Andreas and Lippmann, Judith M. and Varykhalov, Andrei and Duppel, Viola and Lotsch, Bettina V. and Ast, Christian R. and Schoop, Leslie M.},
doi = {10.1088/1367-2630/aa4f65},
issn = {1367-2630},
journal = {New J. Phys.},
month = {dec},
number = {12},
pages = {125014},
publisher = {IOP Publishing},
url = {https://iopscience.iop.org/article/10.1088/1367-2630/aa4f65},
volume = {18},
year = {2016}
}

@article{Moriya1963,
author = {Moriya, T{\^{o}}ru},
doi = {10.1143/JPSJ.18.516},
issn = {0031-9015},
journal = {J. Phys. Soc. Jpn.},
month = {apr},
number = {4},
pages = {516--520},
url = {https://journals.jps.jp/doi/10.1143/JPSJ.18.516},
volume = {18},
year = {1963}
}

@article{Hirose2020,
author = {Hirose, Hishiro T. and Terashima, Taichi and Wada, Taichi and Matsushita, Yoshitaka and Okamoto, Yoshihiko and Takenaka, Koshi and Uji, Shinya},
doi = {10.1103/PhysRevB.101.245104},
issn = {2469-9950},
journal = {Phys. Rev. B},
month = {jun},
number = {24},
pages = {245104},
publisher = {American Physical Society},
url = {https://link.aps.org/doi/10.1103/PhysRevB.101.245104},
volume = {101},
year = {2020}
}

@article{Kawamura2020,
archivePrefix = {arXiv},
arxivId = {2012.08115},
author = {Kawamura, Taiki and Ohki, Daigo and Zhou, Biao and Kobayashi, Akiko and Kobayashi, Akito},
doi = {10.7566/JPSJ.89.074704},
eprint = {2012.08115},
issn = {0031-9015},
journal = {J. Phys. Soc. Jpn.},
month = {jul},
number = {7},
pages = {074704},
url = {https://journals.jps.jp/doi/10.7566/JPSJ.89.074704},
volume = {89},
year = {2020}
}

@article{Zhou2019,
author = {Zhou, Biao and Ishibashi, Shoji and Ishii, Tatsuru and Sekine, Takahiko and Takehara, Ryosuke and Miyagawa, Kazuya and Kanoda, Kazushi and Nishibori, Eiji and Kobayashi, Akiko},
doi = {10.1039/C9CC00218A},
issn = {1359-7345},
journal = {Chem. Commun.},
number = {23},
pages = {3327--3330},
pmid = {30741309},
publisher = {Royal Society of Chemistry},
url = {http://xlink.rsc.org/?DOI=C9CC00218A},
volume = {55},
year = {2019}
}

@article{Kawamura2021,
archivePrefix = {arXiv},
arxivId = {2012.08132},
author = {Kawamura, Taiki and Zhou, Biao and Kobayashi, Akiko and Kobayashi, Akito},
doi = {10.7566/JPSJ.90.064710},
issn = {0031-9015},
journal = {J. Phys. Soc. Jpn.},
month = {jun},
number = {6},
pages = {064710},
url = {https://journals.jps.jp/doi/10.7566/JPSJ.90.064710},
volume = {90},
year = {2021}
}

@article{Tian2021,
author = {Tian, Yefan and Zhu, Yanglin and Li, Rui and Mao, Zhiqiang and Ross, Joseph H.},
doi = {10.1103/PhysRevB.104.L041105},
issn = {2469-9950},
journal = {Phys. Rev. B},
month = {jul},
number = {4},
pages = {L041105},
url = {http://arxiv.org/abs/2104.01311 http://dx.doi.org/10.1103/PhysRevB.104.L041105 https://link.aps.org/doi/10.1103/PhysRevB.104.L041105},
volume = {104},
year = {2021}
}

@article{Barthel1993,
author = {Barthel, E. and Quirion, G. and Wzietek, P. and J{\'{e}}rome, D. and Christensen, J. B. and Jorgensen, M. and Bechgaard, K.},
doi = {10.1209/0295-5075/21/1/015},
issn = {12864854},
journal = {Epl},
number = {1},
pages = {87--92},
volume = {21},
year = {1993}
}

@article{Takane2018,
author = {Takane, Daichi and Souma, Seigo and Nakayama, Kosuke and Nakamura, Takechika and Oinuma, Hikaru and Hori, Kentaro and Horiba, Kouji and Kumigashira, Hiroshi and Kimura, Noriaki and Takahashi, Takashi and Sato, Takafumi},
doi = {10.1103/PhysRevB.98.041105},
issn = {24699969},
journal = {Phys. Rev. B},
number = {4},
pages = {1--5},
publisher = {American Physical Society},
volume = {98},
year = {2018}
}

@article{Hirata2017,
author = {Hirata, Michihiro and Ishikawa, Kyohei and Matsuno, Genki and Kobayashi, Akito and Miyagawa, Kazuya and Tamura, Masafumi and Berthier, Claude and Kanoda, Kazushi},
doi = {10.1126/science.aan5351},
issn = {0036-8075},
journal = {Science},
month = {dec},
number = {6369},
pages = {1403--1406},
pmid = {29242340},
url = {https://www.science.org/doi/10.1126/science.aan5351},
volume = {358},
year = {2017}
}

@article{Kobayashi2021,
author = {Kobayashi, Akiko and Zhou, Biao and Takagi, Rina and Miyagawa, Kazuya and Ishibashi, Shoji and Kobayashi, Akito and Kawamura, Taiki and Nishibori, Eiji and Kanoda, Kazushi},
doi = {10.1246/bcsj.20210230},
issn = {0009-2673},
journal = {Bull. Chem. Soc. Jpn.},
month = {oct},
number = {10},
pages = {2540--2562},
url = {http://www.journal.csj.jp/doi/10.1246/bcsj.20210230},
volume = {94},
year = {2021}
}

@article{Kato2017,
author = {Kato, Reizo and Cui, HengBo and Tsumuraya, Takao and Miyazaki, Tsuyoshi and Suzumura, Yoshikazu},
doi = {10.1021/jacs.6b12187},
issn = {0002-7863},
journal = {J. Am. Chem. Soc.},
month = {feb},
number = {5},
pages = {1770--1773},
pmid = {28121146},
url = {https://pubs.acs.org/doi/10.1021/jacs.6b12187},
volume = {139},
year = {2017}
}

@article{Takane2018a,
author = {Takane, Daichi and Nakayama, Kosuke and Souma, Seigo and Wada, Taichi and Okamoto, Yoshihiko and Takenaka, Koshi and Yamakawa, Youichi and Yamakage, Ai and Mitsuhashi, Taichi and Horiba, Koji and Kumigashira, Hiroshi and Takahashi, Takashi and Sato, Takafumi},
doi = {10.1038/s41535-017-0074-z},
isbn = {4153501700},
issn = {2397-4648},
journal = {npj Quantum Mater.},
month = {dec},
number = {1},
pages = {1},
url = {https://iopscience.iop.org/article/10.1088/1367-2630/aa4f65 http://www.nature.com/articles/s41535-017-0074-z http://arxiv.org/abs/1708.06874 http://dx.doi.org/10.1038/s41535-017-0074-z},
volume = {3},
year = {2018}
}

@article{Tian2020,
author = {Tian, Yefan and Ghassemi, Nader and Ross, Joseph H.},
doi = {10.1103/PhysRevB.102.165149},
issn = {24699969},
journal = {Phys. Rev. B},
month = {oct},
number = {16},
pages = {165149},
publisher = {American Physical Society},
url = {https://link.aps.org/doi/10.1103/PhysRevB.102.165149},
volume = {102},
year = {2020}
}

@article{Dora2009,
archivePrefix = {arXiv},
arxivId = {0904.0116},
author = {D{\'{o}}ra, Bal{\'{a}}zs and Simon, Ferenc},
doi = {10.1103/PhysRevLett.102.197602},
issn = {00319007},
journal = {Phys. Rev. Lett.},
number = {19},
pages = {1--4},
volume = {102},
year = {2009}
}

@article{Kato2017a,
author = {Kato, Reizo and Suzumura, Yoshikazu},
doi = {10.7566/JPSJ.86.064705},
issn = {0031-9015},
journal = {J. Phys. Soc. Jpn.},
month = {jun},
number = {6},
pages = {064705},
url = {http://journals.jps.jp/doi/10.7566/JPSJ.86.064705},
volume = {86},
year = {2017}
}

@article{Kar2020,
author = {Kar, Indrani and Chatterjee, Joydeep and Harnagea, Luminita and Kushnirenko, Y. and Fedorov, A. V. and Shrivastava, Deepika and B{\"{u}}chner, B. and Mahadevan, P. and Thirupathaiah, S.},
doi = {10.1103/PhysRevB.101.165122},
issn = {2469-9950},
journal = {Phys. Rev. B},
month = {apr},
number = {16},
pages = {165122},
publisher = {American Physical Society},
url = {https://doi.org/10.1103/PhysRevB.101.165122 https://link.aps.org/doi/10.1103/PhysRevB.101.165122},
volume = {101},
year = {2020}
}

@article{Yamakage2016,
author = {Yamakage, Ai and Yamakawa, Youichi and Tanaka, Yukio and Okamoto, Yoshihiko},
doi = {10.7566/JPSJ.85.013708},
issn = {0031-9015},
journal = {J. Phys. Soc. Jpn.},
month = {jan},
number = {1},
pages = {013708},
url = {http://journals.jps.jp/doi/10.7566/JPSJ.85.013708},
volume = {85},
year = {2016}
}

@article{Fu2019,
author = {Fu, B.-B. and Yi, C.-J. and Zhang, T.-T. and Caputo, M. and Ma, J.-Z. and Gao, X. and Lv, B. Q. and Kong, L.-Y. and Huang, Y.-B. and Richard, P. and Shi, M and Strocov, V. N. and Fang, C. and Weng, H.-M. and Shi, Y.-G. and Qian, T. and Ding, H.},
doi = {10.1126/sciadv.aau6459},
issn = {2375-2548},
journal = {Sci. Adv.},
month = {may},
number = {5},
pages = {eaau6459},
pmid = {31058219},
url = {https://www.science.org/doi/10.1126/sciadv.aau6459 http://www.ncbi.nlm.nih.gov/pubmed/31058219 http://www.pubmedcentral.nih.gov/articlerender.fcgi?artid=PMC6499591},
volume = {5},
year = {2019}
}

@article{Hirata2021,
author = {Hirata, Michihiro and Kobayashi, Akito and Berthier, Claude and Kanoda, Kazushi},
doi = {10.1088/1361-6633/abc17c},
issn = {13616633},
journal = {Rep. Prog. Phys.},
number = {3},
pages = {036502},
pmid = {33059346},
publisher = {IOP Publishing},
volume = {84},
year = {2021}
}

@article{Imada1998,
author = {Masatoshi Imada and Atsushi Fujimori and Yoshinori Tokura},
doi = {10.1103/revmodphys.70.1039},
issn = {00346861},
issue = {4 PART I},
journal = {Reviews of Modern Physics},
month = {10},
pages = {1039-1263},
volume = {70},
url = {http://www.tandfonline.com/doi/abs/10.1080/09500839008206369 https://link.aps.org/doi/10.1103/RevModPhys.70.1039},
year = {1998}}

@article{Dagotto2005,
author = {Elbio Dagotto},
doi = {10.1126/science.1107559},
issn = {0036-8075},
issue = {5732},
journal = {Science},
month = {7},
pages = {257-262},
volume = {309},
url = {https://www.science.org/doi/10.1126/science.1107559},
year = {2005}
}

@book{Slichter1990,
   author = {Charles P. Slichter},
   city = {Berlin, Heidelberg},
   doi = {10.1007/978-3-662-09441-9},
   isbn = {978-3-642-08069-2},
   publisher = {Springer Berlin Heidelberg},
   title = {Principles of Magnetic Resonance},
   volume = {1},
   url = {http://link.springer.com/10.1007/978-3-662-09441-9},
   year = {1990}
}

@article{Ohki2020,
author = {Ohki, D. and Yoshimi, K. and Kobayashi, A.},
doi = {10.1103/PhysRevB.102.235116},
issn = {24699969},
journal = {Phys. Rev. B},
number = {23},
pages = {1--12},
volume = {102},
year = {2020}
}

@article{Narath1968,
   author = {Albert Narath and H. T. Weaver},
   doi = {10.1103/PhysRev.175.373},
   issn = {0031-899X},
   issue = {2},
   journal = {Physical Review},
   month = {11},
   pages = {373-382},
   volume = {175},
   url = {https://link.aps.org/doi/10.1103/PhysRev.175.373},
   year = {1968}
}

@article{Dressel2007,
   author = {Martin Dressel},
   doi = {10.1007/s00114-007-0227-1},
   issn = {00281042},
   issue = {7},
   journal = {Naturwissenschaften},
   pages = {527-541},
   pmid = {17479236},
   volume = {94},
   year = {2007}
}
\end{document}